\documentclass[twocolumn,superscriptaddress,preprintnumbers,amsmath,amssymb,prl]{revtex4}
\usepackage{graphicx}

\begin{document}

\thispagestyle{empty}

\title{Demonstration of an Unusual Thermal Effect in the Casimir Force from Graphene
}

\author{M. Liu}
\affiliation{Department of Physics and Astronomy, University of California, Riverside, California 92521, USA}

\author{Y. Zhang}
\affiliation{Department of Physics and Astronomy, University of California, Riverside, California 92521, USA}

\author{
G.~L.~Klimchitskaya}
\affiliation{Central Astronomical Observatory at Pulkovo of the
Russian Academy of Sciences, Saint Petersburg,
196140, Russia}
\affiliation{Institute of Physics, Nanotechnology and
Telecommunications, Peter the Great Saint Petersburg
Polytechnic University, Saint Petersburg, 195251, Russia}

\author{
V.~M.~Mostepanenko}
\affiliation{Central Astronomical Observatory at Pulkovo of the
Russian Academy of Sciences, Saint Petersburg,
196140, Russia}
\affiliation{Institute of Physics, Nanotechnology and
Telecommunications, Peter the Great Saint Petersburg
Polytechnic University, Saint Petersburg, 195251, Russia}
\affiliation{Kazan Federal University, Kazan, 420008, Russia}

\author{
 U.~Mohideen\footnote{Umar.Mohideen@ucr.edu}}
\affiliation{Department of Physics and Astronomy, University of California, Riverside, California 92521, USA}

\begin{abstract}
We report precision measurements of the gradient of the Casimir force
between an Au-coated sphere and graphene sheet deposited on a silica plate.
The measurement data are compared with exact theory using the polarization
tensor found in the framework of the Dirac model including effects of the
nonzero chemical potential and energy gap of the graphene sample with no
fitting parameters. The very good agreement between experiment and theory
demonstrates the unusually big thermal effect at separations below $1~\mu$m
which has never been observed for conventional 3D materials. Thus, it is
confirmed experimentally that for graphene the effective temperature is
determined by the Fermi velocity rather than by the speed of light.
\end{abstract}

\maketitle

Over the last two decades special attention has been given to
graphene which is a 2D sheet of carbon atoms packed in
a hexagonal lattice. The  quasiparticles in graphene
are either massless or very light. At energies up to a few eV they are
well described by the relativistic Dirac equation in 2+1 dimensions
where the speed of light $c$ is replaced with the Fermi velocity
$v_F\approx c/300$ \cite{1,2,3}. As a result, graphene offers many
advantages over conventional materials with regard to its mechanical,
electrical, optical, and chemical properties. It possesses the
universal minimum electrical conductivity, low absorbance in the
 range from visible to infrared light and extremely high
mechanical strength \cite{1,2,3}.

One more outstanding feature is an unusually big thermal correction
to the Casimir force between two parallel graphene sheets spaced at
separations below $1~{\mu}$m which was  predicted by
G. G\'{o}mez-Santos \cite{4}. In general, the Casimir force \cite{5}
acts between any two closely spaced surfaces. It is a generalization
of the commonly known van der Waals force \cite{6} taking into account
the finite speed of light. Both forces are of entirely quantum
nature. They are caused by the zero-point and thermal fluctuations of
the electromagnetic field. Since fluctuations have a profound impact
in many physical phenomena, the Casimir effect has gained recognition
as one of the multidisciplinary areas of current research
(see, e.g., the reviews \cite{7,8,9,10,11} and monographs
\cite{11a,12,13,14,15,16,17}).

The Casimir force is described by the fundamental Lifshitz theory
\cite{18,19}. The force value at temperature $T$ can be presented as
the zero-temperature contribution plus a thermal correction.
For two parallel plates made of 3D materials
 spaced at separations  $a<1~{\mu}$m, $T=300~$K
the thermal correction to the Casimir force is  very small
and is not yet measured.
The reason is that
$T\ll T_{\rm eff}=\hbar c/(2ak_B)\approx 1140~$K at $a=1~\mu$m.
It becomes equal to a sizable fraction
of the force only at separations of a few micrometers where the force
itself is too small.  It should be noted that a metal described by
the Drude model is an
exception \cite{20}, but the few percent thermal effect predicted in
this case at $a<1~{\mu}$m was unambiguously excluded
experimentally \cite{21,22}.
There is an extensive literature regarding the reasons for this exclusion
but the problem remains as yet unresolved \cite{8,11,15,22}
(see also a recent overview and a novel approach in Ref.~\cite{22a}).
Quite apart from the case of 3D bodies, according to the prediction
of Ref.~\cite{4}, for two pristine graphene sheets (undoped and with
massless quasiparticles) the thermal correction to the Casimir force
at $T=300~$K becomes large even at separations of tens of nanometers.
This is because for graphene
$T_{\rm eff}^{(g)}=\hbar v_F/(2ak_B)\approx T_{\rm eff}/300$ \cite{4}.
By introducing the thermal photon wavelength
$\lambda_T=2\pi\hbar c/(k_BT)\approx 4.8~\mu$m at $T=300~$K, one finds
that for a pristine graphene the high-temperature regime takes place
at $a\geqslant \lambda_T/300= 16~$nm.

In this Letter, we present the measurement results for the gradient
of the Casimir force between an Au-coated microsphere and a graphene
sheet deposited on a silica glass (SiO${}_2$) plate
obtained in high vacuum using an atomic force microscope (AFM) based
technique operated in the dynamic regime. Although for two parallel 
graphene sheets suspended in a vacuum the thermal correction at short
separations reaches several tens percent of the Casimir force
\cite{4,23}, this configuration is experimentally not feasible. In
the previously performed experiment \cite{24} using a graphene sheet
deposited on a SiO${}_2$ film covering a Si plate, the gradient of
the Casimir force was measured and found in good agreement with
theory \cite{25}. However, the thermal effect could not be discriminated
because the charge carrier concentration in a Si plate was burdened with
a large error and the used theory did not take into account
that real graphene samples are unavoidably doped. In the present
experiment, an improved AFM setup and a much thicker SiO${}_2$
substrate (as suggested in Ref.~\cite{26}) are used. We have also
removed the Si plate and made a comparison with exact
theory taking into account the nonzero chemical potential and energy
gap of the graphene sample. As a result, the measured gradients of the
Casimir force were found to be in a very good agreement with theory
and the thermal effect was reliably demonstrated in the separation
range from 250 to 590~nm where it constitutes from 4\% to 10\% of
the total force gradient.

The Casimir force gradient measurement system consists of a tipless
AFM cantilever \cite{27} whose spring constant was reduced through
chemical etching before use. The corresponding cantilever resonant
frequency was decreased from $5.7579 \times 10^4$ to
$3.5286 \times 10^4~$rad/s by etching for 100~s in 60\% KOH
solution at 75$^{\circ}$C with stirring. The cantilever
was washed in BOE solution and deionized water for 1~min each prior to
etching. The etching process reduced the spring constant of the bare
cantilever as described in Refs.~\cite{40,41}.
 As in previous experiments, a hollow
glass microsphere was next attached to the end of the cantilever
using silver epoxy and then coated with Au. The thickness of the Au
coating and the diameter of the coated sphere were measured to be
$120 \pm 3~$nm and $120.7 \pm 0.1~{\mu}$m using an AFM and a
scanning electron microscope, respectively.  The rms roughness of
the Au coating was measured to be $\delta_s = 0.9 \pm 0.1~$nm. The
resonant frequency of the complete Au coated cantilever-sphere
system was measured in vacuum to be
$\omega_0 = 6.1581 \times 10^3$~rad/s.

The large area graphene monolayer \cite{28} originally chemical vapor
deposition (CVD) grown on a Cu foil was transferred onto a polished JGS2 grade
fused silica double side optically polished substrate of 100~mm
diameter and thickness of $500~{\mu}$m \cite{29} through an
electrochemical delamination procedure \cite{28,30}. A
$1 \times 1~\mbox{cm}^2$ piece of the graphene coated fused
silica wafer was then
cut from the large sample and used in the experiment. The roughness
of the graphene on the fused silica substrate was measured using an
AFM to be $\delta_g = 1.5 \pm 0.1$~nm after the force
gradient measurements.

The graphene impurity concentration was determined after the Casimir
force gradient measurement utilizing Raman spectroscopy, which was
carried out using a Horiba Labram HR 800 system with 532 nm laser
excitation (Laser Quantum, 65~mW power). A 100x objective lens with
NA = 0.9 was used, which leads to a laser spot size of
$\sim 0.4~\mu\mbox{m}^2$ (i.e., of 709~nm diameter). Experiments were
performed at $T = 294 \pm 0.5$~K. A 600~l/m grating was used to
ensure the spectral range of interest (from 1450~cm$^{-1}$ to
2900~cm$^{-1}$) which includes both G and 2D disorder peaks of graphene.
Before collection of spectra at each position, the signal
intensity was maximized in real-time by adjusting the focus of the
microscope, so that graphene sample was located in the focal plane.
The spectrometer calibration was performed as per manual by
reflecting the incident light. Spectral resolution for G peak
identification was measured to be 0.1~cm$^{-1}$. The spectra
collected are integrated results of 10 acquisitions with each
acquisition spanning over 10 s. The spectra were fit to Lorentzians
to identify the G-peak. The G-peak value was compared to reported
G-peak shifts with charge concentration in Ref.~\cite{31} to
identify the impurity concentration in graphene. The latter was
measured at 18 different uniformly distributed locations on the
sample to arrive at an average impurity concentration of
$n = (4.2 \pm 0.3) \times 10^{12}~\mbox{cm}^{-2}$ where the random
and systematic errors are summed to obtain the maximum possible
value of the total error. The dominant impurity chemical type was
Na resulting from the transfer process used \cite{28}. The
respective zero-temperature value of the chemical potential for
our sample is given by \cite{32}
$\mu = \hbar v_F \sqrt{\pi n} = 0.24 \pm 0.01~$eV. This value is
relatively large and, thus, almost independent of $T$ \cite{33}.
Regarding the energy gap $\Delta$, for graphene on a SiO${}_2$
substrate its values vary between 0.01 and 0.2~eV
\cite{34,35,36,37,38}.

Specifics of the vacuum chamber setup have been reported in detail
in previous publications \cite{39,39a,40,41}. The schematic is
provided in Fig.~1 of Ref.~\cite{41}. The fused silica-supported
graphene sample and gold sphere probe were loaded into the vacuum
chamber which was pumped down to below $9 \times 10^{-9}$
Torr using an oil free scroll pump and turbo pump prior to the force
gradient  measurements. Due to the sensitive nature of the graphene
sample, the UV/Ar-ion radiation treatment used in previous
experiments \cite{39a,40,41,41a} for cleaning the Au surfaces was not
implemented to avoid potential damage to the atomic layer graphene.
The cantilever oscillation frequency and motion of the graphene surface
were monitored by two fiber interferometers with 1550~nm and
500.1~nm laser light sources, respectively. A small separation distance
change originating from mechanical drifts during the time span of
the measurements were also taken into consideration. Details of the
above mentioned operations and mechanisms are explained in
Refs.~\cite{39,40,41}. The gradient of the Casimir force was
measured using a dynamic technique, where the force-induced
frequency-shift of cantilever oscillation is directly taken
using a phase lock loop (PLL) \cite{39}. The cantilever
oscillation amplitude was maintained at 10~nm and the PLL
resolution was measured to be 55.3~mrad/s. To ensure the accuracy of
the measurement, the residual potential difference between the gold and
graphene surface was determined through a standard electrostatic
calibration procedure \cite{39,40,41}.

In the dynamic measurement scheme used here, the total force
$F_{\rm tot}(a) = F_{\rm el}(a) + F(a)$ acting on the Au coated sphere,
where $F_{\rm el}(a)$ and $F(a)$ are the electric and Casimir forces,
respectively, and $a$ is the separation distance between the sphere
and graphene, modifies the resonant natural frequency of the
cantilever-sphere oscillator system. The change in the frequency
$\Delta\omega = \omega_r - \omega_0$, where $\omega_r$ is the
resonance frequency in the presence of external force $F_{\rm tot}(a)$,
was recorded by the PLL
every 0.14~nm while the graphene plate was moved towards
the grounded sphere starting at the maximum separation. This was
repeated with each of ten different voltages $V_i$ varied between
0.083 and 0.183~V and eleven voltages equal to the residual
potential difference $V_0$ (see below) applied to the graphene
using ohmic contacts  while the sphere
remained grounded.

The gradients of the total and Casimir forces were found from the
measured frequency shifts using electrostatic calibration. To
perform the calibration of the setup, we used an expression for
the electric force in sphere-plate geometry
$F_{\rm el}(a) = X(a,R)(V_i - V_0)^2$.
Here, $X(a,R)$ is a known function \cite{15,41} and $V_0$ is the
residual potential difference between a sphere  and a
graphene sheet which is nonzero even when both surfaces are
grounded. In the linear regime realized in our
measurement the gradient of the Casimir force is given by \cite{41}
\begin{equation}
F^{\prime}(a) = - \frac{\Delta\omega}{C}-(V_i-V_0)^{2}
\frac{\partial X(a,R)}{\partial a},
\label{eq1}
\end{equation}
\noindent
where $C = \omega_0/(2k)$ and $k$ is the spring constant of the
cantilever. Note that the absolute separations between the zero
levels of the roughness on the sphere and graphene are found from
$a = z_{\rm piezo} + z_0$, where $z_{\rm piezo}$ is the graphene
plate movement due to the piezoelectric actuator and $z_0$ is the
closest approach between the Au sphere and graphene.

{}From the position of the maximum in the parabolic dependence of
$\Delta\omega$ on $V_i$ in Eq.~(\ref{eq1}), one can determine $V_0$
with the help of a $\chi^2$-fitting procedure. From the curvature of
the parabola with the help of the same fit it is possible to determine
$z_0$ and $C$. This was done at all the graphene-Au sphere separations
used in our experiment. In Fig.~1 we present the values of $V_0$ as
a function of separation determined from the fit.
The obtained values were corrected for mechanical drift
of the frequency-shift signal, as discussed in Refs.~\cite{39,40,41}.
As can be seen from Fig.~1, the resulting values of $V_0$ do not
depend on separation. To check this observation, we have performed
the best fit of $V_0$ to the straight lines $V_0 = d + \theta a$
where $a$ is measured in nanometers. It was found that
$d = 0.1326~$V and the slope is $\theta = - 2.73 \times 10^{-7}$~V/nm,
i.e., the independence of
$V_0$ on $a$ was confirmed to a high precision. This finally leads
to the mean value $V_0 = 0.1324$~V.
Next the quantities $z_0$ and
$C$ were determined from the fit at different separations and found
to be separation independent. For our graphene sample the mean
values were found to be $z_0 = 236.9 \pm 0.6$~nm and
$C =(4.599 \pm 0.003) \times 10^5$~s/kg. From the measured resonant
frequency we have confirmed that the obtained value of $C$ results
in the spring constant $k$ consistent with the value determined
prior to the Au coating of the cantilever.

At each separation, the gradient of the Casimir force was measured
21 times with different applied voltages mentioned above. The
mean values of the gradient of the Casimir force were found from
Eq.~(\ref{eq1}) with a step of 1~nm.  The random errors of the mean
were determined at a 67\% confidence level and combined in
quadrature with the systematic errors which are mostly determined
by the errors in measuring the frequency shifts. The obtained
measurement data for $F_{\rm expt}^{\prime}(a)$ with their errors
are shown in Fig.~2 as crosses where $\Delta a = 0.6$~nm [for visual
clarity in Fig.~2(a) all data points are indicated whereas
in Figs.~2(b,c) each second data point and in Fig.~2(d) only each
third one are shown].
\begin{figure}[!t]
\vspace*{-7.7cm}
\centerline{
\includegraphics[width=4.50in]{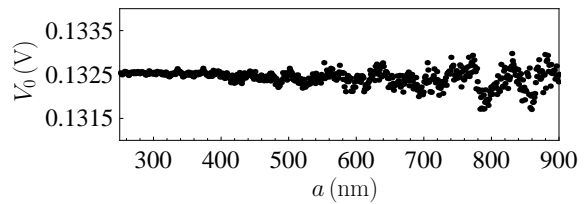}}
\vspace*{-5.8cm}
\caption{\label{fig1} The residual potential difference between a Au-coated sphere
and a graphene sample is shown by the dots as a function of separation. }
\end{figure}

Now we compare the measurement results with theory. In Ref.~\cite{4}, the Casimir
force in graphene systems was calculated using the density-density correlation
functions in the random-phase approximation (see also \cite{42}). This formalism
is equivalent to the nonretarded version of the Lifshitz theory (in so doing,
the relativistic effects were shown to be insignificant \cite{4}). Here, we
use the relativistic version of the Lifshitz formula with reflection coefficients
expressed via the exact polarization tensor of graphene in the framework of the
Dirac model taking into account the
nonzero energy gap $\Delta$ and chemical potential $\mu$ \cite{43,44,45,46}.
Note that even at the shortest separation considered ($a=250~$nm) the characteristic
energy of the Casimir effect $\hbar\omega_c=\hbar c/(2a)=0.4~$eV is well within the
application region of the Dirac model. Because of this the absorption peak of
graphene at $\lambda=270~$nm ($\hbar\omega=2\pi\hbar c/\lambda\approx 4.59~$eV)
does not influence the obtained results.
Then, using the proximity force approximation (PFA) \cite{15}, the gradient
of the Casimir force acting between a Au sphere of radius $R$ and a graphene-coated
SiO$_2$   plate spaced at the separation $a$ at temperature $T$ is given by
\cite{23,25}
\begin{eqnarray}
&&
F^{\prime}(a,T)=2k_BTR\sum_{l=0}^{\infty}{\vphantom{\sum}}^{\prime}
\int_0^{\infty}q_lk_{\bot}dk_{\bot}
\nonumber \\
&&~~
\times\sum_{\alpha}\left[r_{\alpha}^{-1}(i\xi_l,k_{\bot})
R_{\alpha}^{-1}(i\xi_l,k_{\bot})e^{2aq_l}-1\right]^{-1}\!\! .
\label{e1}
\end{eqnarray}

\begin{widetext}
\begin{figure*}[!t]
\vspace*{-5.7cm}
\centerline{
\includegraphics[width=8.20in]{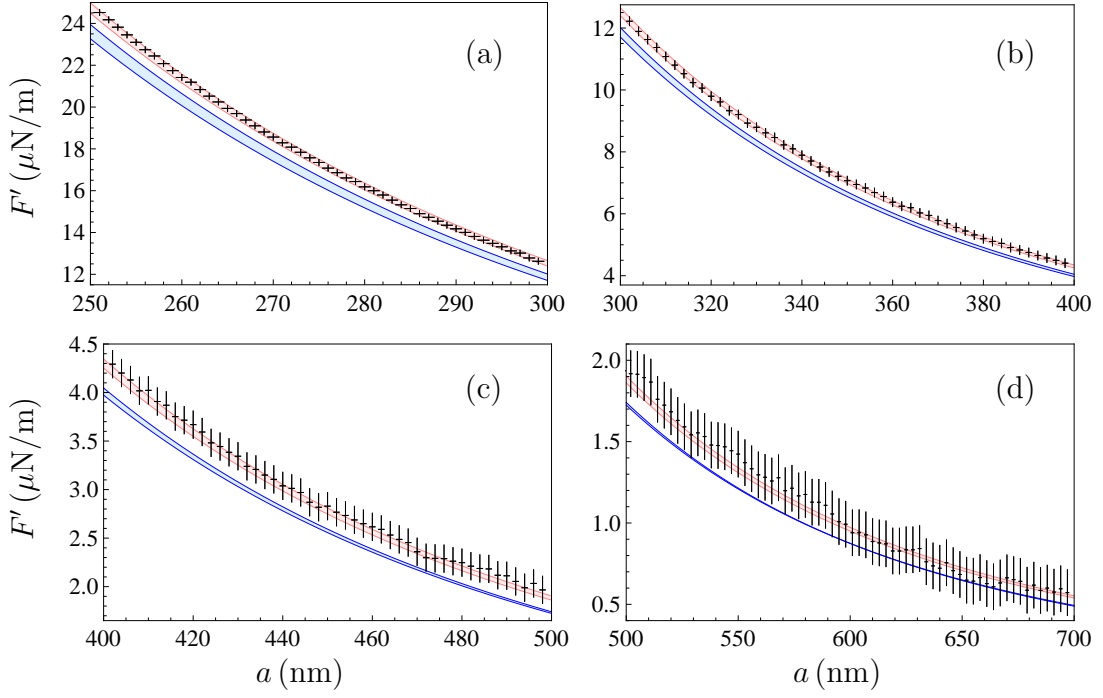}}
\vspace*{-14.8cm}
\caption{\label{fig1} The mean gradient of the Casimir force is shown by the crosses
as a function of separation. The top and bottom theoretical bands
are computed at $T = 294~$K and 0~K, respectively (see the text for
further discussion).  }
\end{figure*}
\end{widetext}

Here,  the prime on the summation sign in $l$
divides the term with $l=0$ by 2,
$k_{\bot}$ is the magnitude of the wave vector projection on the graphene,
$\xi_l=2\pi k_BTl/\hbar$ are the Matsubara frequencies,
$q_l=\sqrt{k_{\bot}^2+{\xi_l^2}/{c^2}}$, and the summation in
$\alpha$ is over the transverse magnetic ($\alpha={\rm TM}$) and
transverse  electric ($\alpha={\rm TE}$) polarizations of the electromagnetic field
(note that the thicknesses of Au coating and SiO$_2$ plate allow consideration of
a sphere as all-gold and a plate as a semispace \cite{15}).
The reflection coefficients $r_{\alpha}$ on the boundaries between Au and vacuum
and $R_{\alpha}$ between vacuum and graphene-coated plate are given by
\cite{23,25,47}
\begin{eqnarray}
&&
r_{\rm TM}(i\xi_l,k_{\bot})=
\frac{\varepsilon_l^{(1)}q_l-k_l^{(1)}}{\varepsilon_l^{(1)}q_l+k_l^{(1)}},
{\ }
r_{\rm TE}(i\xi_l,k_{\bot})=
\frac{q_l-k_l^{(1)}}{q_l+k_l^{(1)}},
\nonumber \\
&&
R_{\rm TM}(i\xi_l,k_{\bot})=
\frac{\hbar k_{\bot}^2(\varepsilon_l^{(2)}q_l-k_l^{(2)})+
q_lk_l^{(2)}\Pi_{00,l}} {\hbar k_{\bot}^2(\varepsilon_l^{(2)}q_l+k_l^{(2)})+
q_lk_l^{(2)}\Pi_{00,l}},
\nonumber\\
&&
R_{\rm TE}(i\xi_l,k_{\bot})=
\frac{\hbar k_{\bot}^2(q_l-k_l^{(2)})-
\Pi_{l}} {\hbar k_{\bot}^2(q_l+k_l^{(2)})+\Pi_{l}},
\label{e2}
\end{eqnarray}
\noindent
where $\varepsilon_l^{(n)}=\varepsilon^{(n)}(i\xi_l)$ with $n=1,\,2$ are the
dielectric permittivities of Au and SiO$_2$, respectively, and
$k_l^{(n)}=\sqrt{k_{\bot}^2+\varepsilon_l^{(n)}{\xi_l^2}/{c^2}}$.
The components of the polarization tensor of graphene are
$\Pi_{\beta\gamma,l}\equiv\Pi_{\beta\gamma}(i\xi_l,k_{\bot},T,\Delta,\mu)$,
where $\beta,\,\gamma=0,\,1,\,2$, and the combination of these components
$\Pi_l$ is defined as
\begin{equation}
\Pi_l=k_{\bot}^2\Pi_{\beta,\,l}^{\,\,\beta}-q_l^2\Pi_{00,l}.
\label{e3}
\end{equation}

It is convenient to present the quantities $\Pi_{00,l}$ and $\Pi_l$ in the
form
\begin{equation}
\Pi_{00,l}=\Pi_{00,l}^{(0)}+\Pi_{00,l}^{(1)},
\quad
\Pi_{l}=\Pi_{l}^{(0)}+\Pi_{l}^{(1)},
\label{e4}
\end{equation}
\noindent
where the first contributions describe undoped graphene ($\mu=0$) with a nonzero
energy gap ($\Delta=2mv_F^2$, $m$ is the mass of quasiparticles) at $T=0$ but
with $\omega=i\xi_l$,
and the second ones take into account an explicit dependence of the polarization
tensor on $T$ and $\mu$. It has been shown that \cite{43,44}
\begin{equation}
\Pi_{00,l}^{(0)}=\frac{\alpha\hbar k_{\bot}^2}{\tilde{q}_l}\Psi(D_l),
\quad
\Pi_{l}^{(0)}=\alpha\hbar\tilde{q}_l\Psi(D_l),
\label{e5}
\end{equation}
\noindent
where $\tilde{q}_l=\sqrt{v_F^2k_{\bot}^2+\xi_l^2}/c$,
$\alpha=e^2/(\hbar c)$ is the fine structure constant,
$\Psi(x)=2[x+(1-x^2)\arctan(x^{-1})]$, and $D_l=\Delta/(\hbar c\tilde{q}_l)$.

The exact expressions for $\Pi_{00,l}^{(1)}$ and $\Pi_l^{(1)}$ can be found
in Refs.~\cite{46,47}
\begin{eqnarray}
&&
\Pi_{00,l}^{(1)}=\frac{4\alpha\hbar c^2\tilde{q}_l}{v_F^2}\int_{D_l}^{\infty}du
\left(\sum_{\kappa=\pm 1}\frac{1}{e^{B_lu+\kappa\frac{\mu}{k_BT}}+1}
\right)
\nonumber \\
&&~~
\times\left[1-{\rm Re}\frac{1-u^2+2i\gamma_lu}{(1-u^2+2i\gamma_lu+D_l^2-
\gamma_l^2D_l^2)^{1/2}}\right],
\nonumber \\
&&
\Pi_{l}^{(1)}=-\frac{4\alpha\hbar \tilde{q}_l\xi_l^2}{v_F^2}\int_{D_l}^{\infty}du
\left(\sum_{\kappa=\pm 1}\frac{1}{e^{B_lu+\kappa\frac{\mu}{k_BT}}+1}
\right)
\nonumber \\
&&~~
\times\left[1-{\rm Re}\frac{(1+i\gamma_l^{-1}u)^2+(\gamma_l^{-2}-1)D_l^2}{(1-u^2
+2i\gamma_lu+D_l^2-\gamma_l^2D_l^2)^{1/2}}\right],
\label{e6}
\end{eqnarray}
\noindent
where $\gamma_l\equiv\xi_l/(c\tilde{q}_l)$ and $B_l\equiv\hbar c\tilde{q}_l/(2k_BT)$.

The values of $\varepsilon_l^{(n)}$ where obtained \cite{8,15} by means of the
Kramers-Kronig relation using the tabulated optical data for Au and SiO$_2$ \cite{48}.
Note that due to the smallness of the reflection coefficient $R_{\rm TE}(0,k_{\bot})$
the values of force gradients are almost independent of the type of
extrapolation of the available optical data for Au down to zero frequency.

The gradients of the Casimir force computed by Eqs.~(\ref{e1})--(\ref{e6}) at $T=294~$K
were corrected (a fraction of 1\% effect) for the presence of surface roughness \cite{8,15}
\begin{equation}
F_{\rm theor}^{\prime}(a,T)=\left(1+10\frac{\delta_s^2+\delta_g^2}{a^2}\right)
F^{\prime}(a,T).
\label{e7}
\end{equation}

The same computations have been repeated at $T=0$ when the summation
on $l$ in Eq.~(\ref{e1})
is replaced with integration (the explicit expressions for
$\Pi_{00}(i\xi,k_{\bot},0,\Delta,\mu)$ and  $\Pi(i\xi,k_{\bot},0,\Delta,\mu)$ are
contained in Ref.~\cite{47}).

The computational results for the boundaries of allowed theoretical bands are shown in
Fig.~2 by the top and bottom pairs of lines computed at $T=294~$K and $T=0~$K,
respectively. These lines were obtained in the following most conservative manner.
The upper lines in both pairs were computed for $\mu =0.25~$eV, $\Delta=0~$eV,
whereas the lower lines --- for  $\mu =0.23~$eV, $\Delta=0.2~$eV (we recall that with
increasing $\mu$ and $\Delta$ the force gradient increases and decreases, respectively).
Keeping in mind that  the PFA slightly increases the force gradients,
we did not correct the upper lines for the PFA errors but introduced
the maximum possible correction factor of $(1 - a/R)$ to the lower
boundary lines \cite{49,50,51,52,53}. The widths of theoretical bands
have also been increased to incorporate errors in the sphere radius
and optical data of Au and SiO${}_2$.

As is seen in Fig. 2, the measurement data are in excellent
agreement with theoretical predictions at $T = 294$K. The unusual
thermal effect in the force gradient equal to the difference between
the top and bottom bands is conclusively demonstrated over the
region from 250 to 590 nm. Specifically, at
$a$ = 250, 300, 400, 500, and 590 nm the thermal correction reaches
4\%, 5\%, 7\%, 8.5\%, and 10\% of the total force gradient,
respectively.
This correction is smaller than for a pristine graphene because it is suppressed
by the relatively large value of $\mu$. The thermal correction is contributed
by a summation over the $T$-dependent Matsubara frequencies and an explicit
dependence of the polarization tensor on $T$ as a parameter. For a pristine
graphene both effects contribute almost equally \cite{23}. In our case a summation
over the Matsubara frequencies contributes 70\%, 81\%, and 88\% of the thermal
correction at $a=250$, 400, and 600~nm, respectively.

To conclude, we demonstrated the unusual thermal effect in the Casimir force
from graphene at separations below $1~\mu$m. Although similar effects
are the subject of considerable literature, they have never been
observed in measurements of the Casimir interaction at short
separations. This result is important not only for the
fundamental investigations of graphene and for Casimir physics, but
for numerous applications in nanoscale science.

The work of M.~L., Y.~Z.~and U.~M.~was partially supported by the NSF
grant PHY-2012201.
The work of G.~L.~K. and V.~M.~M. was partially supported by the Peter
the Great Saint Petersburg Polytechnic
University in the framework of the Russian state assignment for basic
research (project N FSEG-2020-0024).
V.~M.~M.~was partially funded by the Russian Foundation for Basic
Research, Grant No. 19-02-00453 A. His work was also partially
supported by the Russian Government Program of Competitive Growth
of Kazan Federal University.

\end{document}